\documentclass[twocolumn,aps,superscriptaddress,nofootinbib,floatfix]{revtex4}
\usepackage{amsfonts}
\usepackage{float}
\usepackage{placeins}
\usepackage{graphicx}
\usepackage[hidelinks]{hyperref}
\usepackage{color,amsmath,amssymb,bm}
\usepackage{tensor}

\usepackage[normalem]{ulem}  

\renewcommand{\sout}{\bgroup \color{red} \ULdepth=-.5ex \ULset}

\def\blfootnote{\xdef\@thefnmark{}\@footnotetext}

\newcommand{\beq}{\begin{equation}}
\newcommand{\eeq}{\end{equation}}
\newcommand{\bea}{\begin{eqnarray}}
\newcommand{\eea}{\end{eqnarray}}

\begin{document}

\title{The signature of charge dependent directed flow observables by electromagnetic fields in heavy ion collisions}

\author{Yifeng Sun}
\email{sunyfphy@lns.infn.it}
\affiliation{Laboratori Nazionali del Sud, INFN-LNS, Via S. Sofia 62, I-95123 Catania, Italy}
\affiliation{Department of Physics and Astronomy, University of Catania, Via S. Sofia 64, 1-95125 Catania, Italy}

\author{Vincenzo Greco}
\email{greco@lns.infn.it}
\affiliation{Laboratori Nazionali del Sud, INFN-LNS, Via S. Sofia 62, I-95123 Catania, Italy}
\affiliation{Department of Physics and Astronomy, University of Catania, Via S. Sofia 64, 1-95125 Catania, Italy}

\author{Salvatore Plumari}
\email{salvatore.plumari@ct.infn.it}
\affiliation{Department of Physics and Astronomy, University of Catania, Via S. Sofia 64, 1-95125 Catania, Italy}
\affiliation{Laboratori Nazionali del Sud, INFN-LNS, Via S. Sofia 62, I-95123 Catania, Italy}

\date{\today}

\begin{abstract}
We discuss the generation of the directed flow $v_1(p_T,y_z)$ induced by the electromagnetic field as a function of $p_T$ and $y_z$.
Despite the complex dynamics of charged particles due to strong interactions generating several anisotropies in the azimuthal 
angle, it is possible at $p_T > m$ to directly correlate the splitting in $v_1$ of heavy quarks with different charges to some main features of the magnetic field,
and in particular its values at formation and freeze-out time.

We further found that the slope of the splitting $d\Delta v_1/dy_z|_{y_z=0}$ of  positively and negatively charged particles at high $p_T$ can be formulated as $d\Delta v_1/dy_z|_{y_z=0}=-\alpha \frac{\partial \ln f}{\partial p_T}+\frac{2\alpha-\beta}{p_T}$, where $f$ is the $p_T$ spectra of the charged particles and the constants $\alpha$ and $\beta$ (order of MeV) are constrained by the $y$ component of magnetic fields and the sign of $\alpha$ is simply determined by the difference $\Delta[tB_y(t)]$ in the center of colliding systems at the formation time of particles and at the time when particles leave the effective range of electromagnetic fields or freeze out. 
The formula is derived from general considerations and is confirmed by several related numerical simulations; it 
supplies a useful guide to quantify the effect of different magnetic field configurations and provides an evidence 
of why the measurement of $\Delta v_1$ of charm, bottom and leptons from $Z^0$ decay and their correlations 
are a powerful probe of  the initial e.m. fields in ultra-relativistic collisions.

\end{abstract}

\maketitle

\section{Introduction}

The ultrarelativistic heavy ion collisions (uRHICs) at both the BNL Relativistic Heavy Ion Collider (RHIC)~\cite{Adams:2005dq,Adcox:2004mh} and the CERN Large Hadron Collider (LHC)~\cite{Aamodt:2008zz} 
have created a new state of matter, the quark-gluon plasma (QGP),  during their early stage and showed that such a matter is the most perfect fluid in nature \cite{Kovtun:2004de,Romatschke:2007mq,Gale:2013da}. In the last decades, there are numerous studies in the search for the parity ($P$) and charge conjugate parity ($CP$) symmetry breaking processes in quantum chromodynamics (QCD) happened in QGP, mainly via the chiral magnetic effect (CME)~\cite{Kharzeev:2007jp,Fukushima:2008xe,Kharzeev:2009fn,Jiang:2016wve,Shi:2017cpu,Sun:2018idn} 
and the chiral vortical effect (CVE)~\cite{Sun:2018onk}. The strongest ever electromagnetic (e.m.) field and the largest relativistic vorticity \cite{Becattini:2017gcx,STAR:2017ckg} are created in non-central heavy ion collisions.
The strong e.m. field can lead to many other interesting phenomena such as the chiral magnetic wave (CMW)
~\cite{Kharzeev:2010gd,Burnier:2011bf,Yee:2013cya,Sun:2016nig} and the splitting in the spin polarization of hyperons
~\cite{Becattini:2016gvu,Han:2017hdi,Guo:2019joy}. 
However, there are a lot of uncertainties in the calculation of magnetic field  in heavy ion collisions, and this inspired the search for a direct probe to the strong e.m. fields by measuring the charge dependent flows ($v_n$) of charged mesons and baryons 
as well as neutral charged charmed mesons~\cite{Gursoy:2014aka,Das:2016cwd}.
Though these numerical studies 
 are very meaningful for the understanding of e.m. fields, a general description of charge dependent flows
going beyond the details of e.m. fields is also important. In Refs~\cite{Sun:2020wkg}, we have found several important features of the directed flow splitting, $\Delta v_1$, of positively and negatively charged heavy quarks as well as leptons induced by e.m. fields: (i) It is not very sensitive to the details of the spatial and time configurations of e.m. fields; (ii) $d\Delta v_1/d\eta$ at mid-pseudorapidity ($\eta$) depends on the slope of the transverse momenta  spectra, especially at high $p_T$; (iii) The modification by the interaction with QGP for heavy quarks  is negligible at $p_T$ higher than 2-3 GeV$/c$. This persuades us to find the general physics behind these features and extend the findings to other charge dependent flow observables. Since the effect due to the interaction with QGP is small for charged quarks at high $p_T$, the findings should have a general application for high $p_T$ heavy quarks and energetic jets as well as leptons of arbitrary $p_T$.  The purpose of these studies is not only trying to build the bridge between the spatial and temporal configurations of e.m. fields and final observables in the theoretical side, but also can be used to determine whether the $\Delta v_n$ observed experimentally has an electromagnetic origin.

The configuration of non-central heavy ion collisions is fixed throughout the paper, where the center of the nuclei moving in positive $z$ direction localizes in positive $x$ axis, which produces a strong magnetic field and large vorticity in negative $y$ direction. When mentioning the formula in this paper, 
it should be noted that it is deduced 
by assuming a pure interaction with e.m. fields, which means that it should be applied to high $p_T$ heavy quarks and to leptons 
at arbitrary $p_T$ where the strong interaction with QGP leads to negligible modifications. 
The numerical results from transport simulation for heavy quarks include, however, the strong interaction with QGP according to the current
standard approach to their dynamics.

The paper is organized as follows: In Sec. II, we describe the general formula for the charge dependent flow observables induced by e.m. fields, and give the physical meaning to the coefficients in the general formula. In Sec. III, we study specifically the  $\Delta v_1$ of positively and negatively charged particles, and present the direct connection between the coefficients in $\Delta v_1$ and the $y$ component of the magnetic field. Sec. IV presents several numerical results that can be understood by the general formula probing the robustness of the formula. We also discuss the importance of measuring the $\Delta v_1$
and propose also a new measurement of the spectra ratio of positively and negatively charged leptons from $Z^0$ decay. 
Summary and conclusions are discussed in Sec. V.

\section{Charge dependent flow harmonics by electromagnetic fields}

The effects of the e.m. fields on the phase-space distribution function can be expressed in terms of a transition function $T(\Delta p_x,\Delta p_y,\Delta y_z,p_x,p_y,y_z)$. The function $T$ represents the distribution of the shifts in the transverse momenta $\mathbf{p}_T= (p_x,p_y)$ and rapidity $y_z$
due to the electromagnetic field. 
In order to guarantee the particle number conservation the distribution function $T$ satisfy the following normalization condition $\int d^2\Delta p_T d\Delta y_z T(\Delta p_x,\Delta p_y,\Delta y_z,p_x,p_y,y_z) = 1$. 
Starting from a boost-invariant spectra of charged particles $f(p_T)$, after the modification by e.m. fields, which is 
considered as a small perturbation, the distribution $f^{'}(\mathbf{p}_T,y_z)$ shall be:
\begin{eqnarray}
&&f^{'}(\mathbf{p}_T, y_z) =\int d^2\Delta p_T d\Delta y_z f(\mathbf{p}_T-\Delta \mathbf{p}_T,y_z-\Delta y_z)\nonumber
\\&& \times T(\Delta \mathbf{p}_T,\Delta y_z,\mathbf{p}_T-\Delta \mathbf{p}_T,y_z-\Delta y_z)\nonumber
\\&& \approx \int d^2\Delta p_T d\Delta y_z [ f(\mathbf{p}_T,y_z)T(\Delta \mathbf{p}_T,\Delta y_z,\mathbf{p}_T,y_z)\nonumber
\\&&- \frac{\partial f T}{\partial p_x}\Delta p_x
- \frac{\partial fT}{\partial p_y} \Delta p_y\nonumber
-\frac{\partial fT}{\partial y_z} \Delta y_z ] \nonumber
\\&=&f-(\frac{\partial f\overline{\Delta p}_x}{\partial p_x}+\frac{\partial f\overline{\Delta p}_y}{\partial p_y}+f\frac{\partial \overline{\Delta y}_z}{\partial y_z}),
\end{eqnarray}
where $f T= f(\mathbf{p}_T,y_z)T(\Delta \mathbf{p}_T,\Delta y_z ,\mathbf{p}_T,y_z)$ and the derivatives are evaluated at $(\mathbf{p}_T,y_z)$,
and the average shifts $\overline{\Delta p}_a$ with $a=x,y,z$ are defined as:
\begin{eqnarray}
\overline{\Delta p}_a (\mathbf{p}_T,y_z)&=&\int d^2\Delta p_T d\Delta y_z \, T(\Delta \mathbf{p}_T,\Delta y_z,\mathbf{p}_T,y_z) \, \Delta p_a .\nonumber
\\&&
\end{eqnarray}
By the definition of rapidity, one can further express $\overline{\Delta y}_z$ in terms of $\overline{\Delta p}_x$, $\overline{\Delta p}_y$, $\overline{\Delta p}_z$ as:
\begin{eqnarray}
\overline{\Delta y}_z&=&-\frac{p_T \tanh{y_z}}{m_T^2}(\cos{\phi} \overline{\Delta p}_x+\sin \phi \overline{\Delta p}_y)\nonumber
\\&+&\frac{\overline{\Delta p}_z}{m_T \cosh y_z},
\end{eqnarray}
where $\phi=\tan^{-1}(p_x/p_y)$ is the azimuthal angle relative to the reaction plane  in momentum space.
Since the colliding systems are symmetric with $y\leftrightarrow-y$, in momentum space one should have  $\overline{\Delta p}_x(p_T,\phi,y_z)=\overline{\Delta p}_x(p_T,2\pi-\phi,y_z)$, $-\overline{\Delta p}_y(p_T,\phi,y_z)=\overline{\Delta p}_y(p_T,2\pi-\phi,y_z)$ and $\overline{\Delta p}_z(p_T,\phi,y_z)=\overline{\Delta p}_z(p_T,2\pi-\phi,y_z)$. Therefore, after a Fourier decomposition with respect to the angle $\phi$, the average shift can be expressed as:
\begin{eqnarray}
\overline{\Delta p}_x&=&\sum 2a_n(p_T,y_z) \cos n\phi,\nonumber
\\\overline{\Delta p}_y&=&\sum 2b_n(p_T,y_z) \sin n\phi, \nonumber
\\\overline{\Delta p}_z&=&\sum 2c_n(p_T,y_z) \cos n\phi. \label{shifts}
\end{eqnarray}
In heavy ion collisions, if the chiral  magnetic conductivity is zero, there is no $B_z$~\cite{PhysRevC.94.044903,Inghirami:2019mkc}, and according to Lorentz force the momentum shifts are:
\begin{eqnarray}
\overline{\Delta p}_x&=&q\int dt (E_x-v_z B_y),\nonumber
\\\overline{\Delta p}_y&=&q\int dt (E_y+v_z B_x), \nonumber
\\\overline{\Delta p}_z&=&q\int dt (E_z+v_xB_y-v_yB_x).\label{force1}
 \end{eqnarray}
The above set of equations suggests that the coefficients $a_n, b_n$ and $c_n$ have direct relations with the e.m. fields as a function of $\phi_x$ in coordinate space.
Since 
\begin{eqnarray}
\frac{\partial}{\partial p_x}=\cos \phi\frac{\partial}{\partial p_T}-\frac{\sin \phi}{p_T}\frac{\partial}{\partial \phi}\nonumber
\\\frac{\partial}{\partial p_y}=\sin \phi\frac{\partial}{\partial p_T}+\frac{\cos \phi}{p_T}\frac{\partial}{\partial \phi},
 \end{eqnarray}
 then the distribution function $f^{'}(\mathbf{p}_T,y_z)$ after the modification of e.m. fields relates to the initial $f(\mathbf{p}_T,y_z)$ as:
\begin{eqnarray}
f^{'}&=&f-\{\frac{\partial f(a_1+b_1)}{\partial p_T}+f(-\frac{p_T}{m_T^2}\frac{\partial (a_1+b_1)\tanh y_z}{\partial y_z}\nonumber
\\&+&\frac{a_1+b_1}{p_T}+\frac{2}{m_T}\frac{\partial c_0/\cosh y_z}{\partial y_z})\}\nonumber
\\&-&\lbrace -f\frac{p_T}{m_T^2}\frac{\partial (a_0+b_0)\tanh y_z}{\partial y_z}+\frac{\partial (a_0+b_0)f}{\partial p_T} \rbrace \cos \phi\nonumber
\\&-&\sum_{n=1} \{\frac{\partial f(a_{n+1}+b_{n+1}+a_{n-1}-b_{n-1})}{\partial p_T}\nonumber
\\&+&f[\frac{(n+1)(a_{n+1}+b_{n+1})-(n-1)(a_{n-1}-b_{n-1})}{p_T} \nonumber
\\&-&\frac{p_T}{m_T^2}\frac{\partial\tanh y_z(a_{n+1}+b_{n+1}+a_{n-1}-b_{n-1})}{\partial y_z}\nonumber
\\&+&\frac{2}{m_T}\frac{\partial c_n/\cosh y_z}{\partial y_z}]\}\cos n\phi.
\label{distribution}
\end{eqnarray}
One can read from Eq. (\ref{distribution}) that the Lorentz force  in the longitudinal direction ($c_n$) leads also to non-zero charge dependent $v_n$ that measures the anisotropy in transverse momenta. This is an effect that was not brought to light in the previous studies focused on the numerical simulations
\cite{Das:2016cwd,Sun:2020hvb,Sun:2020wkg,Oliva:2020doe,Gursoy:2014aka,Gursoy:2018yai}, even if they naturally include it.

If $p_T$ is larger than the mass of charged particles, Lorentz force is not sensitive to $p_T$ any more, because $p_T/m_T \approx 1$
and furthermore the equations of motion do not depend on the $p_T$ of the particle implying similar trajectories. 
All of this together lead to:
\begin{eqnarray}
&&\frac{\partial a_n}{\partial p_T}  \simeq 0, \frac{\partial b_n}{\partial p_T}  \simeq 0, \frac{\partial c_n}{\partial p_T}  \simeq 0 (p_T\gg m),
\label{approx}
\end{eqnarray}
while in general $a_n,b_n, c_n$ are $p_T$ dependent and the specific dependence is determined by the way the strong interaction
with the QGP medium acts on the specific particle. Given that such momentum dependence can be discarded in the high $p_T$ limit,
one can rearrange the terms in Eq.(\ref{distribution}) that can be rewritten as:
\begin{eqnarray}
&&f^{'}=f-\sum_{n=0} (d_n\frac{\partial f}{\partial p_T}+e_n \frac{f}{p_T})\cos n\phi  \,\,\,\,\,(p_T\gg m),\label{d2}
\label{flow}
\end{eqnarray}
where $d_n$ and $e_n$ are mixed combinations of $a_n, b_n$ and $c_n$ which can be read from Eq. (\ref{distribution}). 
Moreover, if two types of charged particles have similar formation time, such as charm quarks and leptons from $Z^0$ decay, then  at $p_T\gg m$ all their coefficients $a_n$, $b_n$ and $c_n$ are differed only by their charges regardless of the complex spatial and temporal configurations of e.m. fields, which provides a strong correlation between their flow observables supplying strong probes of the e.m. fields. One immediate consequence of Eq. (\ref{d2}) is that for charged particles with a peculiar spectra, for example a sudden change in $p_T$ like the leptons from $Z^0$ decay, as long as these coefficients are non-zero, there will be a sudden change in the spectra ratio and the $\Delta v_n$ of positively and negatively charged particles just at exactly the same  $p_T$.

Eq. (\ref{distribution}) provides  a general $p_T$ scaling for all flow observables induced by e.m. fields, and it reduces to Eq. (\ref{d2}) at $p_T\gg m$, for any configuration of e.m. fields. This scaling is certainly different from the collective flows generated by the strong interaction with QGP for light and heavy quarks, since $a_n, b_n$ and $c_n$ induced by e.m. fields are charge dependent, and become constant at $p_T\gg m$, while they are not charge dependent but flavor dependent hence with a $p_T$ dependence determined by the flavor dependence of the strong interactions. 
In a preliminary study presented in a recent Proceedings ~\cite{Sun:2020hvb}, we showed that the interaction with QGP is negligible for the charge 
dependent flow observables induced by e.m. fields at high $p_T$ for  heavy quarks and leptons of arbitrary $p_T$, 
hence the $\Delta v_n$ should be a general signature of effects induced by e.m. fields.
In Section IV the comparison to realistic simulations including the effective hot QCD matter interaction will allow to assess the range of validity 
of the approximations done to deduce Eq.(\ref{d2}).

\section{Charge dependent directed flow induced by electromagnetic fields}
Reading from Eq. (\ref{distribution}), under the conditions in Eq.(\ref{approx}), the directed flow $v_1$ becomes, up to quadrupole moments:
\begin{eqnarray}
v_1&=& \frac{p_T}{m_T^2}\frac{\partial }{\partial y_z} [(a_0 +\frac{1}{2}(a_2+b_2))\tanh y_z]-\frac{1}{m_T}\frac{\partial c_1/\cosh y_z}{\partial y_z}\nonumber
\\&-&[ a_0+ \frac{1}{2}(a_2+b_2)] \frac{\partial \ln f}{\partial p_T}
- \frac{(a_2+b_2)}{p_T}
\end{eqnarray}
In AA collisions in the overlapping region, one finds the quadrupole moment of e.m. fields is smaller than their 0-th order moment, which 
can be seen by looking at the e.m. fields at the very initial stages of AA collisions. With this assumption, one has the simplified:
\begin{eqnarray}
v_1\approx -a_0 \frac{\partial \ln f}{\partial p_T}+\frac{p_T}{m_T^2}\frac{\partial a_0 \tanh y_z}{\partial y_z}-\frac{1}{m_T}\frac{\partial c_1 {\rm sech} y_z}{\partial y_z}
\label{v1}
\end{eqnarray}
It is noted that $v_1$ has a contribution also from the Lorentz force in longitudinal direction. In AA collisions, since the colliding systems are also symmetric with $x,y,z\leftrightarrow-x,-y,-z$, which leads to $a_n(y_z)=(-1)^{n+1}a_n(-y_z)$, $b_n(y_z)=(-1)^{n+1}b_n(-y_z)$ and $c_n(y_z)=(-1)^{n+1}c_n(-y_z)$, so all terms in Eq. (\ref{v1}) are non-zero and $v_1$ is odd in rapidity.
As shown in Eq.(\ref{v1}) at leading order the $v_1$ depends only on the two coefficients $a_0$ and $c_1$.
In order to understand the role of the e.m. field in the building up of the $v_1$, in this section, we study the relation 
between these coefficients and the e.m. field.\\
Starting from Eq. (\ref{force1}) relating the $\overline{\Delta p}_x$ to the time integral of the Lorentz force,
we project into the 0-th order in the azimuthal angle  both sides, which gives for the coefficient $a_0$:
\begin{eqnarray}
 2a_0&=&q\int dt  \{E_{x0}(\rho,t,\eta_s)-\tanh y_z B_{y0}(\rho,t,\eta_s)\}\nonumber
 \\&\approx& q\int dt \{E_{x0}(\rho,t,y_z)-\tanh y_z B_{y0}(\rho,t,y_z)\},\nonumber
 \\
 \label{a00}
 \end{eqnarray}
where we have used transverse coordinates $\rho=\sqrt{x^2+y^2}$ and $\phi_x=\tan^{-1} (x/y)$ while $\eta_s$ is the space-time rapidity $\eta_s=\frac{1}{2}\ln \frac{t+z}{t-z}$. 
In the second line, we make use of the approximation $\eta_s \approx y_z$ valid in a boost invariant geometry. 
Notice that in Eq. (\ref{a00}) $E_{x0}$ and $B_{y0}$ are the 0-th order in a Fourier decomposition of the e.m. field with 
$E_{x0}=\int \frac{d\phi_x}{2\pi} E_x(\rho,\phi_x)$ and $B_{y0}= \int \frac{d\phi_x}{2\pi} B_y(\rho,\phi_x)$.
Though in  Eq. (\ref{a00}) one has to take into account the coordinate  and momentum distributions of charged particles initially produced in the overlap region as well as their corresponding trajectories, it is possible to reduce their complexity.
As shown in Ref~\cite{Sun:2020wkg}, if the e.m. fields do not change too abruptly with respect to $\rho$ in the overlapping region, which is the case in AA collisions, it is possible to show that  the overall effect can be approximated as:
 
\begin{eqnarray} 
&&2a_0(p_T,y_z)\approx qK\int_{t_0}^{\infty}dt \Theta(1-\gamma/R)\nonumber
\\&&\times \{E_x(t,y_z)-\tanh y_z B_y(t,y_z)\}|_{\rho=0}
\label{a0}
\end{eqnarray}
where $K$ is a positive constant 
that depends on the spatial distribution of e.m. fields and the spatial distribution
 of charged particles when they are initially formed, $\Theta$ is the step function, $R$ is a radius about the average of the effective ranges of e.m. fields and overlapping region of colliding systems, and $\gamma$ is  $\frac{p_T}{m_T}(\frac{t}{\cosh y_z}-\tau_0)$ with $t_0=\tau_0\cosh y_z$ the formation time of charged particles. Because of the Faraday's Law, $E_x$ and $B_y$ at $\rho=0$ are related; we express them as:
\begin{eqnarray}
B_y|_{\rho=0}\equiv-g(t,\eta_s),\,\,\,\,\,\,
E_x|_{\rho=0}\equiv h(t,\eta_s),
\end{eqnarray}
  (the negative sign in $B_y$ because its direction stays along the negative $y$ direction). 
  
The understanding of the key features of the strength and time dependence of the electromagnetic field that determines the magnitude and the sign
of $\Delta v_1$, is a main aim of the present work. Our strategy has been to reduce the Maxwell equations to a 1D integral equation
that can give a quite good approximation of the relations between $B_y(t,\eta_s)$  and $E_x(t,\eta_s)$; this is obtained from  
the Faraday's Law $\boldsymbol{\nabla}\times\mathbf{E}=-\partial \mathbf{B}/\partial t$
under the assumption of small space gradients $\frac{\partial E_z}{\partial x}\sim 0$, hence in the inner part of the QGP
fireball this allows to write at $\rho=0$ and small $\eta_s$:
\begin{eqnarray}
h(t,\eta_s)&\approx&h(t,0)+\int_0^{\eta_s} d\chi \frac{t}{\cosh^2\chi}
\left(\frac{\partial g}{\partial t}-\frac{\partial g}{\partial \chi}\frac{\sinh 2\chi}{2t}\right)\nonumber
\\
\label{ex}
\end{eqnarray}
with $h(t,0)=0$, i.e. at the collision center $E_x = 0$.
\begin{figure}[h]
\centering
\includegraphics[width=1\linewidth]{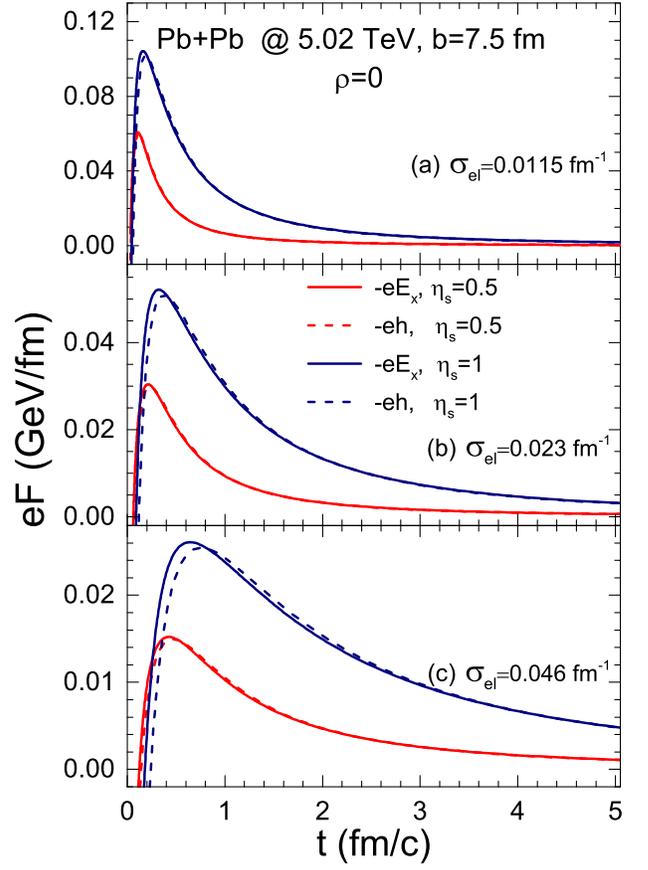}
\caption{(Color online)  The comparison of $E_x$ at $\rho=0$ between the calculations by Eq. (\ref{ex}) neglecting space gradients $\frac{\partial E_z}{\partial x}$  from the time evolution of $B_y$ and that by solving full Maxwell equations with space gradients for three different choices of electrical conductivity $\sigma_{el}$.}
\label{fig:Ex}
\end{figure}

To check the reliability of Eq. (\ref{ex}), we compare $E_x$ at $\rho=0$ given by $h(t,\eta_s)$ through Eq. (\ref{ex}), 
once $B_y(t,\eta_s)=-g(t,\eta_s)$ has been calculated by solving Maxwell equations, with the $E_x$ solution of the full Maxwell equations
that includes space gradients. 
To get an analytical solution of Maxwell equations coupled with conducting medium is quite though in heavy ion collisions, 
and it is still not yet possible to have reliable solutions in magnetohydrodynamics unless in infinite conductivity case~\cite{Inghirami:2019mkc}. 
However, assuming a constant and uniform conductivity, one can obtain the analytical solutions of e.m. fields
~\cite{Tuchin:2013apa,Gursoy:2014aka,PhysRevC.94.044903}, which are adopted by several studies recently
~\cite{Gursoy:2014aka,Das:2016cwd,Gursoy:2018yai,Chatterjee:2018lsx}. 
The analytical results in this case of $E_x$ at $\rho=0$ in 5.02 TeV Pb+Pb collisions at impact parameter $b=7.5$ fm are shown by the solid lines in Fig.~\ref{fig:Ex}, where the different choices of electrical conductivity $\sigma_{el}=$ 0.0115, 0.023 and 0.046 fm$^{-1}$ are included. It is seen that with larger conductivity, the maximum magnitude of $E_x$ decreases, while the lifetime of it increases. 
The dashed lines are $E_x$ calculated by Eq. (\ref{ex}), i.e. $h(t, \eta_s)$ , from the time evolution of $B_y$ at $\rho=0$, that is  given by the analytical results in the same conditions. It is seen that Eq. (\ref{ex}) coincides with analytical results at $\eta_s=0.5$, and it is a good approximation even at $\eta_s=1$. 

By combining Eqs. (\ref{a0}) and (\ref{ex}), one can relate $a_0$ directly to $B_y$ and its time derivative implicitly
including the effect of $E_x$:
\begin{eqnarray}
\frac{2a_0}{qK}&=&\int_{t_0}^{t_f}dt \left[ h(t,y_z)+\tanh y_z g(t,y_z)\right]\nonumber
\\&=&\int_{t_0}^{t_f}dt [ \int_0^{y_z} d\chi \frac{t}{\cosh^2\chi} \left(\frac{\partial g}{\partial t}-\frac{\partial g}{\partial \chi}\frac{\sinh 2\chi}{2t}\right)\nonumber
\\&+& g \tanh y_z ]
=\int_{t_0}^{t_f}dt \int_0^{y_z}d\chi [\frac{t\frac{\partial g}{\partial t}}{\cosh^2\chi}-\tanh \chi \frac{\partial g}{\partial \chi}\nonumber
\\&+&\frac{\partial g\tanh\chi}{\partial \chi}]=\int_{t_0}^{t_f}dt \int_0^{y_z}d\chi\frac{t\frac{\partial g}{\partial t}+g}{\cosh^2\chi}\nonumber
\\&=& \int_0^{y_z}d\chi \frac{1}{\cosh^2\chi} \left[t_fg(t_f,\chi)-t_0g(t_0,\chi)\right],
\label{a0f}
\end{eqnarray}
with $t_0=\tau_0\cosh y_z$ and $ t_f(p_T)=(\tau_0+Rm_T/p_T)\cosh y_z$. The slope $\frac{d\Delta a_0}{dy_z}|_{y_z=0}$ of positively and negatively charged particles and anti-particles is an important quantity of $v_1$ and it becomes the following simple form:
\begin{eqnarray}
\frac{d\Delta a_0}{dy_z}|_{y_z=0}=|q|K\left[\tau_1 g(\tau_1,0)-\tau_0 g(\tau_0,0)\right]\nonumber\\
\simeq -|q|K\ \left[\tau_1 B_y(\tau_1,0)-\tau_0 B_y(\tau_0,0) \right],\label{slope}
\end{eqnarray}
with $\tau_1(p_T)=\tau_0+Rm_T/p_T$. Eq. (\ref{slope}) shows that the sign and magnitude of $\frac{d\Delta a_0}{dy_z}|_{y_z=0}$ is just determined by the difference of  $tB_y$ in the center of fireball at the formation time of particles and the time when particles escape the control of e.m. fields, or when particles freeze out. The detailed information of e.m. fields is irrelevant, of course this comes out under the approximation that the gradients of the fields
are not too large within the inner part of the fireball and the balance between electric and magnetic field scales in a self-similar way with
$r-$space coordinates.
The factorization found in terms of only the $t$ dependence of $B_y$ allows a new insight into the generation of the splitting in
$v_1$ between particles with different charges reducing the delicate balance between the magnetic Lorentz force and the Faraday's effect
in terms of the time evolution of the magnetic field only.

From Eq. (\ref{force1}), since the system is symmetric with $y\leftrightarrow-y$, which leads to a $B_x$ with no 0-th order expansion in $\phi_x$, 
we understand the leading term to $c_1$ from $B_x$ comes from its 2nd order expansion. 
However, this is expected to be small in the overlapping region of colliding systems. The same happens to the 1st order expansion in $\phi_x$ of $E_z$, see  analytical solutions using constant and uniform conductivity in Ref~\cite{PhysRevC.94.044903}. 
The dominant contribution to $c_1$ is thus from $B_y \simeq -g(t, y_z)$ in its 0th order expansion, which can be approximated simply as:
\begin{eqnarray}
2c_1&\approx&-qK\int_{t_0}^{t_f}dt \frac{p_T}{m_T\cosh y_z}g(t,y_z)\nonumber
\\&=&\frac{-qKp_T}{m_T}\int_{\tau_0}^{\tau_1(p_T)}d\tau g(\tau\cosh y_z,y_z).\label{c1f}
\end{eqnarray}
From Eqs.  (\ref{a0f}) and (\ref{c1f}) one finds $a_0$ and $c_1$ are $p_T$ independent at $p_T\gg m$, 
and this holds for any configurations of e.m. fields as we discuss before.  Eqs. (\ref{v1}), (\ref{a0f}) and (\ref{c1f}) capture essential ingredients for the charge dependent $v_1$ induced by e.m. fields and tell what information we can extract from the experimental measurement. 

\section{numerical results}
In this section we will use some specific cases to see how Eqs. (\ref{v1}), (\ref{a0f}) and (\ref{c1f})  can help us better understand numerical results
from realistic simulations. We still choose the 5.02 TeV Pb+Pb collision systems at $b=7.5$ fm and the e.m. fields are given by the analytical solutions of Maxwell equations assuming a constant and uniform conductivity. It should be noted that there is a discontinuity of e.m. fields in the initial stages of heavy ion collisions, since non-zero conductivity has to appear after collision rather than even before. So the numerical study is just a platform to test our analytical formula rather than making predictions to be tested in experiments. The general analytical formula however can be applied to all charge dependent flow observables induced by e.m. fields in AA systems in relativistic heavy ion collisions.

\begin{figure}[h]
\centering
\includegraphics[width=1\linewidth]{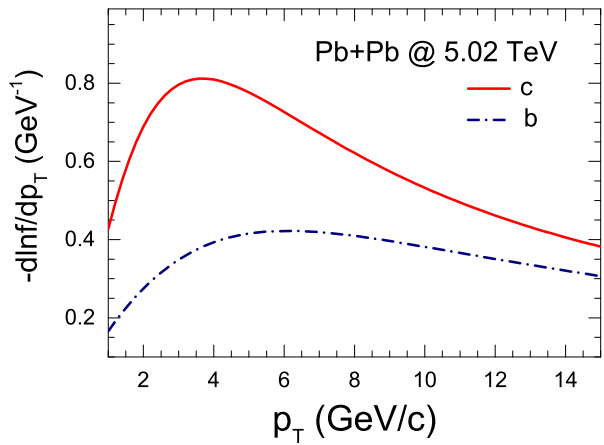}
\caption{(Color online)  $-\frac{\partial \ln f}{\partial p_T}$ of charm and bottom quarks in 5.02 TeV Pb+Pb collisions.
}
\label{fig:cbspectra}
\end{figure}

We include realistic initialization of heavy quarks in transverse momentum space.
Charm quarks are formed at same $\tau_0=0.1$ fm$/c$  and we initialize the momentum distribution of charm 
quarks spectra $f_c$  in 5.02 TeV Pb+Pb collisions with the prompt distribution obtained within
the Fixed Order+Next-to-Leading Log (FONLL) QCD~\cite{PhysRevLett.95.122001,Cacciari:2012ny},
that reproduces the D-mesons spectra in pp collisions after fragmentation.
We parametrize it as:
\begin{eqnarray}
&&f_c=\frac{A}{(1+B {p_T}^{n})^{\alpha}},
\end{eqnarray}
where the parameters are $A=20.28$, $n=1.951$, $\alpha=3.137$ and $B=0.0752$ respectively. 
The solid red line in Fig.~\ref{fig:cbspectra} shows the $p_T$ dependence of $-\frac{\partial \ln f_c}{\partial p_T}$, which is seen 
to approach the maximum value of 0.8 GeV$^{-1}$ at 3.5 GeV$/c$ for charm and 0.4 GeV$^{-1}$ at about 6 GeV$/c$ for bottom.

To study quantitatively the dynamics of HQs we solve the relativistic Langevin equation in an expanding QGP background.
The background medium is described by the relativistic transport code 
with fixed shear viscosity to entropy density ratio close to the lower bound $1/4\pi$ which was constrained by the experimental data on the collective flows of charged particles \cite{Ruggieri:2013ova,Plumari:2015cfa,Plumari:2019gwq,Sun:2019gxg}.
The dynamics of heavy quarks is studied by standard Langevin equations
\cite{Gossiaux:2008jv,Gossiaux:2009mk,Cao:2015hia,Xu:2017obm,PhysRevC.73.034913,PhysRevLett.100.192301,PhysRevC.86.014903,Alberico:2011zy,Alberico:2013bza}
with the inclusion of Lorentz force~\cite{Das:2016cwd,Chatterjee:2018lsx}:
\begin{eqnarray}
&&d x_i = \frac{p_i}{E}dt,
\label{velocity}\\
&&{d p_i} = -\Gamma p_idt+\xi_{i}\sqrt{2D_pdt}+q(E_i+\epsilon_{ijk}v_jB_k)dt,\nonumber
\\
\label{force}
\end{eqnarray}
where the momentum diffusion coefficient $D_p$ is related to the drag coefficient $\Gamma$ by $D_p=\Gamma ET$, and $\xi_i$ is a real number randomly sampled from a normal distribution with $\langle\xi_i\rangle=0$ and $\langle \xi_i\xi_j\rangle=\delta_{ij}$. 
Before the formation of QGP at about 0.3 fm$/c$ in 5.02 TeV Pb+Pb collisions, $\Gamma$ and $D_p$ are set to zero and heavy quarks interact with only e.m. fields. $\Gamma$ and  $D_p$ are derived from  a quasi-particle model (QPM)~\cite{Das:2012ck,Berrehrah:2013mua,Berrehrah:2014kba}. The QPM approach accounts for the non-perturbative dynamics by $T-$dependent quasi-particle masses, with $m_q^2=1/3g^2(T)T^2$ and $m_g^2=3/4g^2(T)T^2$, plus a $T-$dependent background field known as a bag constant,
with $g(T)$ tuned to fit the thermodynamics of the lattice QCD~\cite{Borsanyi:2010cj,Plumari:2011mk}.
This approach has been shown to lead to a good description of the experimental data for both $R_{AA}(p_T)$ and $v_2$ of charmed and bottomed mesons,
both at RHIC and LHC employing  an enhancement factor $K \sim 2$ of the drag and diffusion coefficients~\cite{Scardina:2017ipo,Sun:2019fud,Dong:2019unq}.

\subsection{Transverse momentum dependence of $d\Delta v_1/dy_z$}

\begin{figure}[h]
\centering
\includegraphics[width=1\linewidth]{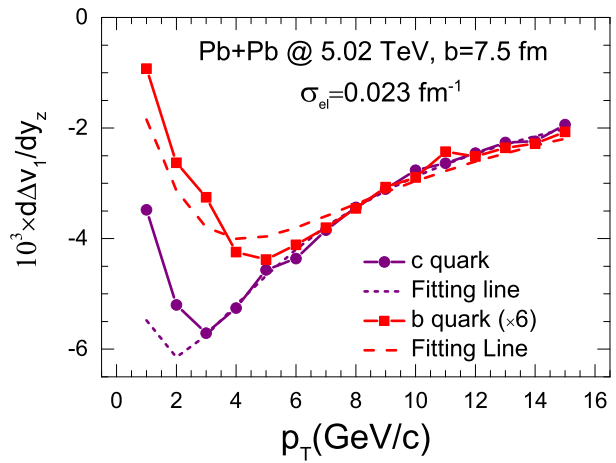}
\caption{(Color online)  The comparison between $d\Delta v_1^c/dy_z|_{y_z=0}$ of  charm and bottom quarks and anti-quarks from numerical simulation and the fitting with $-\alpha\frac{\partial \ln f_c}{\partial p_T}+(2\alpha-\beta)\frac{p_T}{m_T^2}$.
}
\label{fig:dv1dy}
\end{figure}

In this section we present our results on the transverse momentum dependence of the directed flow.
We first studied charm quarks under e.m. fields generated by conducting medium with an electrical conductivity $\sigma_{el}=0.023$ fm$^{-1}$, which is within the bound of LQCD results~\cite{Ding:2010ga,Amato:2013naa,Brandt:2012jc}.
In Fig.~\ref{fig:dv1dy} by the solid purple line, we show the slope $d\Delta v_1^c/dy_z|_{y_z=0}$ of  charm quarks and anti-quarks after the evolution in e.m. fields as well as in QGP. The slope is seen to be negative though very small, and its magnitude increases with $p_T$ initially but then decreases at $p_T>$3 GeV$/c$. The results can be understood exploiting Eqs. (\ref{v1}), (\ref{slope}) and (\ref{c1f}), which lead to the following scaling assuming $\frac{\partial a_0}{\partial p_T}$ close to zero:
\begin{eqnarray}
\frac{d\Delta v_1^c}{dy_z}|_{y_z=0}&=&\frac{d\Delta a_0}{dy_z}|_{y_z=0}
\left(-\frac{\partial \ln f_c}{\partial p_T}+\frac{2p_T}{m_T^2}\right)-\beta\frac{p_T}{m_T^2}\nonumber
\\&=&-\alpha\frac{\partial \ln f_c}{\partial p_T}+(2\alpha-\beta)\frac{p_T}{m_T^2},\label{scaling}
\end{eqnarray}
with
\begin{eqnarray}
\left\{\begin{matrix}
\alpha=|q|K\{\tau_1g(\tau_1,0)-\tau_0g(\tau_0,0)\},\\
\beta=|q|K(\lambda-d^2\lambda/dy_z^2)|_{y_z=0},\\
\lambda(y_z)=\int_{\tau_0}^{\tau_1(p_T)}d\tau g(\tau\cosh y_z,y_z).
\end{matrix}\right.\label{coeff}
\end{eqnarray}
where we recall that the function $g(t,y_z)= -B_y(t,y_z)$.
We see that if $B_y$ does not strongly change with $\eta_s$, $\beta$ is positive.

We used the scaling above to fit the numerical results at $p_T>3$ GeV and found that  the scaling with $\alpha=-3.6$ MeV and $\beta=2.5$ MeV agrees quite well with the numerical results. Moreover, with $\tau_1g(\tau_1,0)\approx 2.5$ MeV, $\tau_0g (\tau_0,0)=11.5$ MeV and $|q_c|=2/3$, one can find $K\approx$0.6. At low $p_T$, there appears a deviation from the scaling, which is mainly due to the suppression by the strong interactions with the QGP~\cite{Sun:2020hvb}.
Still it is remarkable it works also for bottom at $p_T> 6 \, \rm Gev/c$.

\subsection{The effect of conductivity on $d\Delta v_1/dy_z$}

\begin{figure}[h]
\includegraphics[width=1\linewidth]{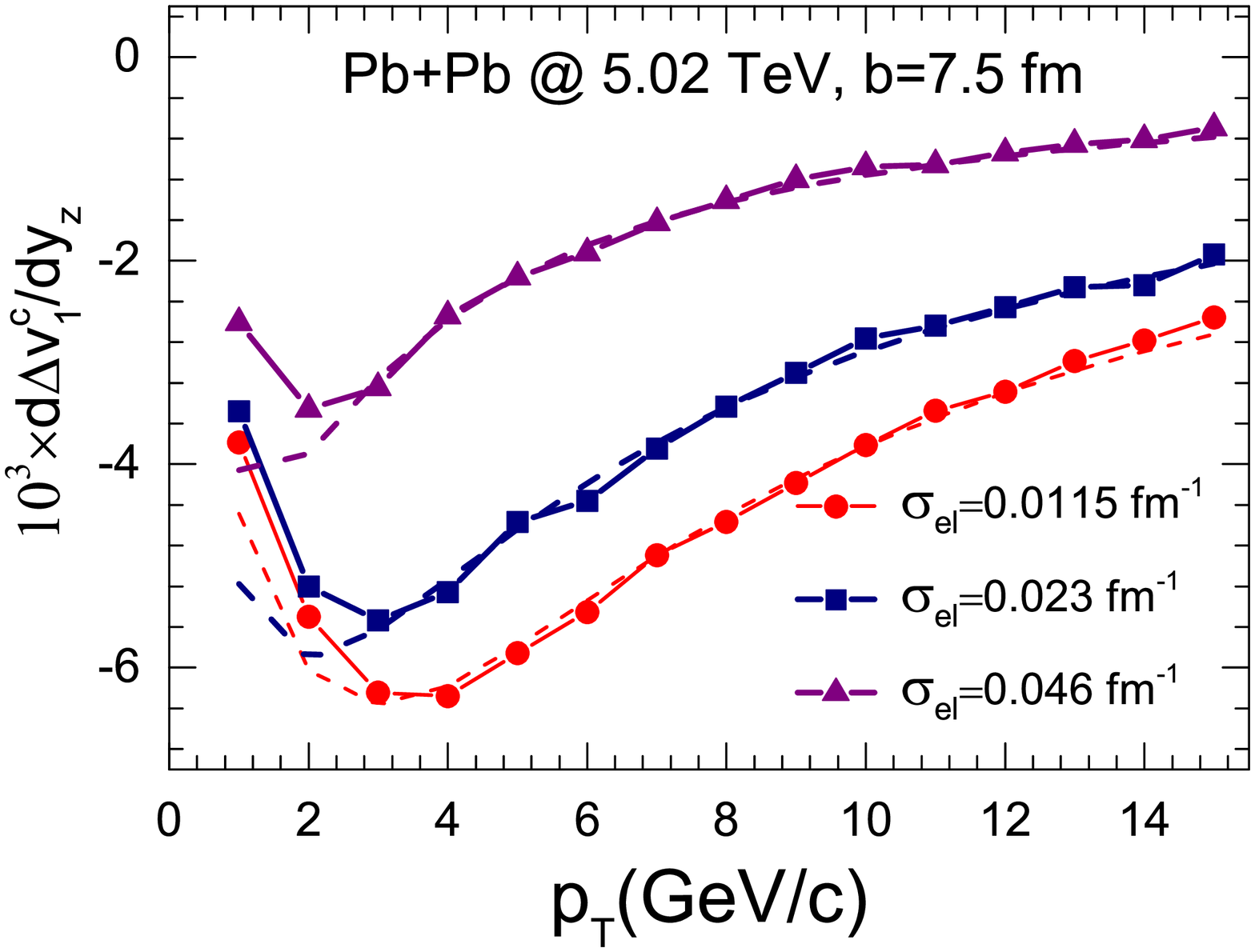}
\includegraphics[width=1\linewidth]{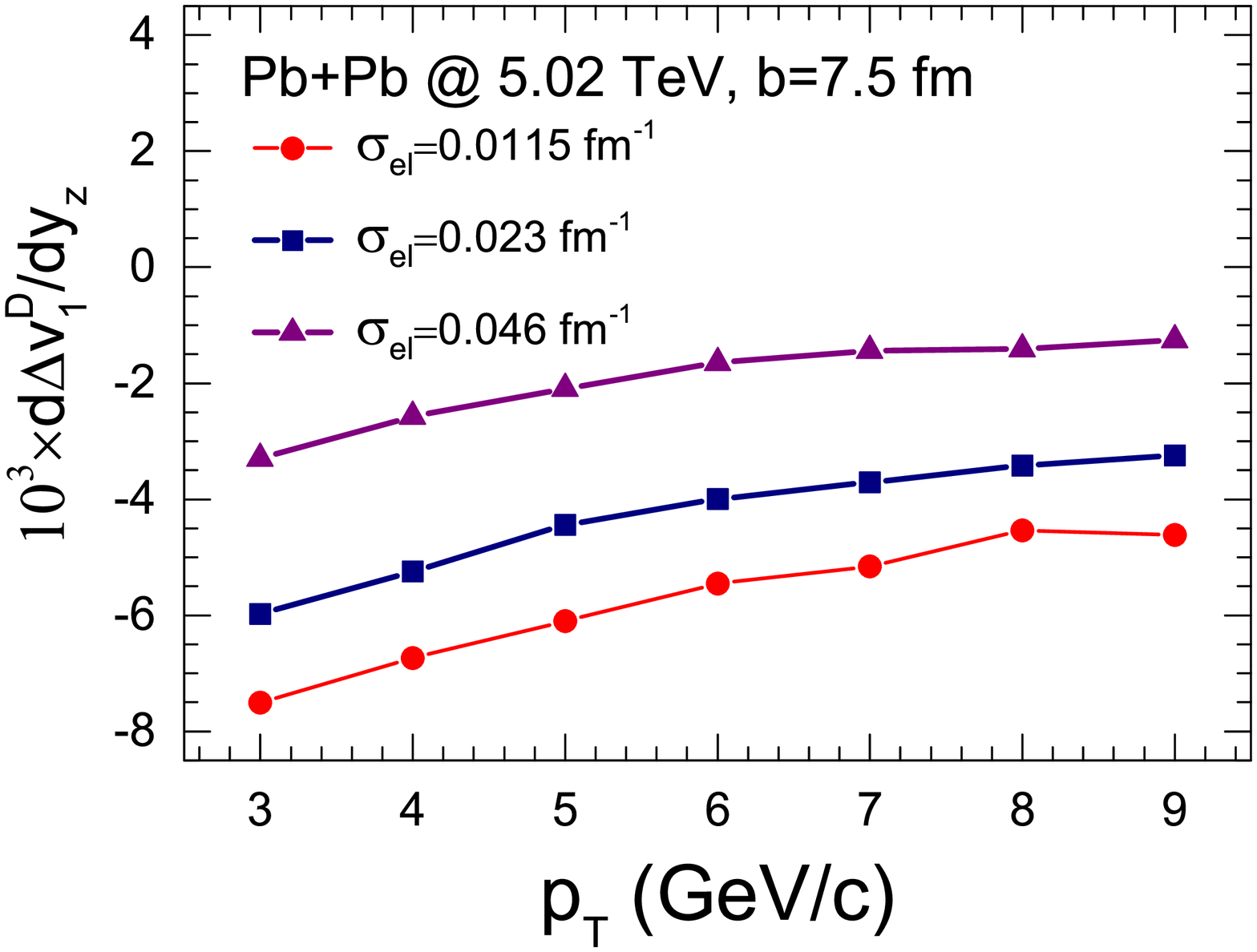}
\caption{(Color online)  $d\Delta v_1/dy_z|_{y_z=0}$ generated by e.m. fields with the medium conductivity $\sigma_{el}=$ 0.0115, 0.023 and 0.46 fm$^{-1}$ in 5.02 TeV Pb+Pb collisions at $b=$ 7.5 fm. Upper panel: symbols are the results for charm(anti-charm quarks), the dashed lines are the fit by Eq. (\ref{scaling}); lower panel: results for $D^0$  ($\rm c\overline{u}$) and $\overline{D}^0$ ($\rm \overline{c}u$).
}
\label{fig:charm}
\end{figure}

\begin{figure}[h]
\centering
\includegraphics[width=1\linewidth]{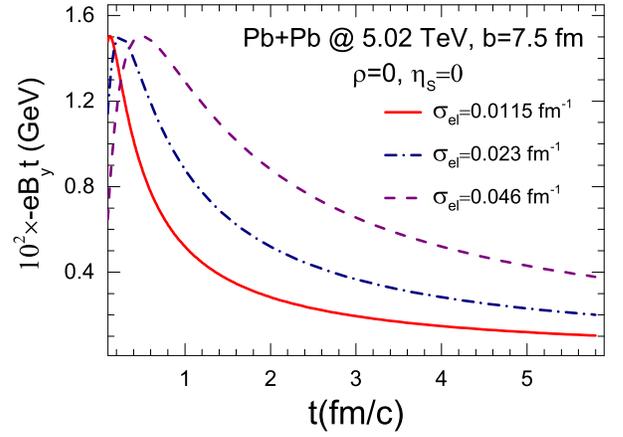}
\caption{(Color online) The product of time and $B_y$ at the center of the colliding systems with different values of electrical conductivity in 5.02 TeV Pb+Pb collisions at $b=$ 7.5 fm.
}
\label{fig:tBy}
\end{figure}

One of key properties of QGP is  its electrical conductivity, and increasing the conductivity can increase the lifetime of e.m. fields, which affects the charge dependent flow observables.  In Fig.~\ref{fig:charm}, we show the variation of $d\Delta v_1^D/dy_z|_{y_z=0}$ of $D^0$  ($\rm c\overline{u}$) and $\overline{D}^0$ ($\rm \overline{c}u$) with the variation of $\sigma_{el}$, where charmed meson are formed by the Peterson fragmentation as done in \cite{Scardina:2017ipo,PhysRevLett.100.192301}.  It is  seen in Fig.~\ref{fig:charm} that with larger conductivity, the magnitude of the  slope becomes smaller. Reading from Eq. (\ref{coeff}), which relates  $\alpha$ directly to  $\tau_1g(\tau_1)-\tau_0g(\tau_0)= -(\tau_1 B_y(\tau_1)-\tau_0 B_y(\tau_0)) $, one can see why it is so
by looking at the evolution of $-tB_y$ at the center of the colliding systems, as shown in Fig.~\ref{fig:tBy}. It is seen that increasing $\sigma_{el}$ 
will decrease the magnitude of magnetic field initially while increase it at latter time. This leads to the decrease of the magnitude of $\alpha$ due to 
the decrease in the difference of $tB_y$ at $\tau_1$ and $\tau_0$. Reading from Fig.~\ref{fig:tBy}, we can obtain $\alpha_1$, $\alpha_2$ and $\alpha_3$ from 0.0115 to 0.46 fm$^{-1}$ taking the ratio $(\alpha_1-\alpha_2):(\alpha_2-\alpha_3)\approx 0.7$, which agrees quite well with the results in Fig.~\ref{fig:charm}.

We have to note that such a result is non trivial because one would expect that a larger conductivity inducing a $B_y$ with a longer
lifetime and a larger strength for nearly all the time evolution, see Fig. \ref{fig:tBy}, would generate a stronger charge/anti-charge splitting of the $v_1$.
This important aspect is caught by the formula we have derived in the previous Section. The physical reason for this behavior
can be understood considering that for small conductivity the quick variation of the magnetic field generates a strong electric field 
by the Faraday's law that wins over the Lorentz magnetic force that acts in the opposite direction.  At increasing conductivity
the magnetic field has a slower evolution thus inducing a smaller electric field and this reduces $v_1$ because there is a nearly exact cancellation
between the magnetic field and the electric field. Under the approximation done we have been able to trace back such
a delicate dynamics in terms of the variation of the $t B_y(t)$ and the slope of the particel spectrum according to
Eqs. (\ref{scaling}) and (\ref{coeff}).
It can be expected that with larger conductivity, $\tau_1g(\tau_1)-\tau_0g(\tau_0)$ becomes positive, and it should lead 
to a positive $d\Delta v_1/dy_z|_{y_z=0}$ that would agree with the experimental measurement in ALICE~\cite{Acharya:2019ijj}; this 
indeed has been seen in Ref. \cite{Sun:2020wkg}.

\subsection{The $\Delta v_1$ from charm to bottom quarks}
Switching from charm to bottom quarks in the study of $\Delta v_n$, one encounters the differences in quark's charge, in the interaction strength with QGP, 
in the initial spectra, in the mass and in the formation time. As the effect by the difference in charge is trivial, and the interaction strength difference plays a negligible role at high $p_T$, we thus focus on the variations of $\Delta v_n$ by the other three differences
whose impact can be envisaged by our factorized formula in Eq. (\ref{scaling}).
To isolate the effects induced by these three differences separately, we make only one change each time. The colliding system is still 5.02 TeV Pb+Pb collisions at $b=7.5$ fm with a fixed electrical conductivity $\sigma_{el}=0.023$ fm$^{-1}$.

\begin{figure}[h]
\centering
\includegraphics[width=1\linewidth]{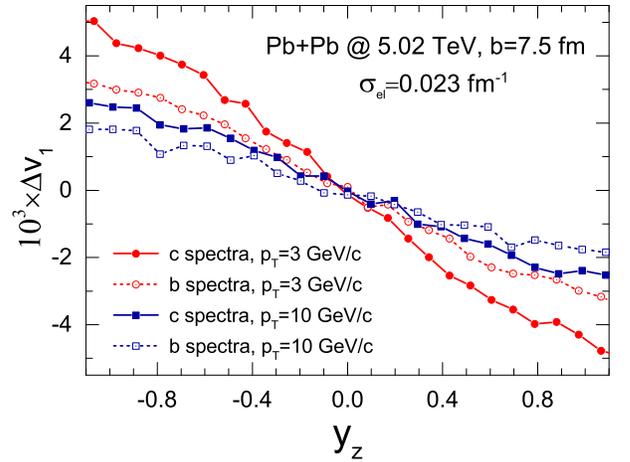}
\caption{(Color online)  The comparison between $\Delta v_1$ vs $y_z$ of mesons fragmented by heavy quarks with the initial spectra taken from charm quarks and from bottom quarks at $p_T=$ 3 and 10 GeV$/c$.
}
\label{fig:cb}
\end{figure}
We first study the effect of particle's spectra on $\Delta v_1$ by using the initial spectra to bottom quarks, which is obtained by FONLL as well. The parameters are found to be $A=0.468$, $n=1.838$, $\alpha=3.076$ and $B=0.0302$ respectively, and the $-\frac{\partial\ln f_b}{\partial p_T}$ is shown by the blue dash-dotted line in Fig.~\ref{fig:cbspectra}.

As shown in Fig.~\ref{fig:cb}, where the slopes $d\Delta v_1/dy_z|_{y_z=0}$ at $p_T=3$ GeV$/c$ are found to be -5.8$\times10^{-3}$ with charm spectra and -3.1$\times10^{-3}$ with bottom spectra and -2.9$\times10^{-3}$ with charm spectra and -2.0$\times10^{-3}$ with bottom spectra at $p_T=10$ GeV$/c$, the spectra taken from bottom quarks decreases the magnitude of the slope of $\Delta v_1$ vs $y_z$, and the suppression is smaller at  $p_T=10$ GeV$/c$ compared to that at $p_T=3$ GeV$/c$. The results can be understood by the general scaling in Eq. (\ref{scaling}), which is generated by their difference in $-\frac{\partial\ln f}{\partial p_T}$ of charm and bottom quarks. Specifically, as shown in Fig.~\ref{fig:cbspectra}, $-\frac{\partial\ln f}{\partial p_T}$ of bottom quarks is always smaller than charm quarks, and their ratio approaches maximum of about a factor of two at $p_T=$3-4 GeV$/c$, while decreasing at high $p_T$.

\begin{figure}[h]
\centering
\includegraphics[width=1\linewidth]{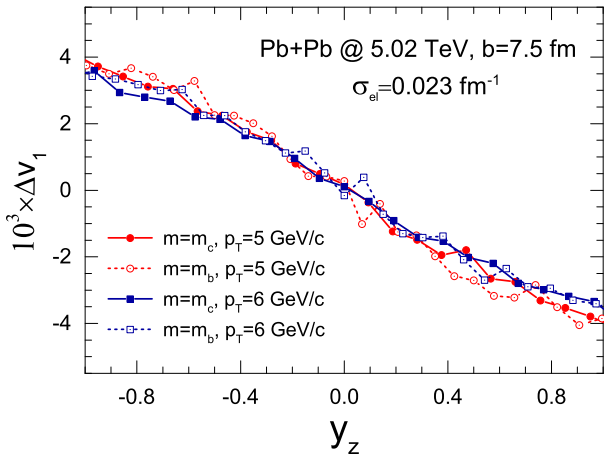}
\caption{(Color online) The comparison between $\Delta v_1$ vs $y_z$ of  mesons fragmented by heavy quarks with the quark mass taken from charm quarks and from bottom quarks at $p_T=$ 5 and 6 GeV$/c$.
}
\label{fig:cbmass}
\end{figure}
In Fig.~\ref{fig:cbmass}, we study the effect of  mass of quarks on $\Delta v_1$ vs $y_z$ of  mesons fragmented by heavy quarks, where we solely change the mass of quarks from $m_c=1.3$ GeV to $m_b=4$ GeV, and the mesons from $m_D=1.87$ GeV to $m_B=5.27$ GeV. As shown in Fig.~\ref{fig:cbmass}, the slopes $d\Delta v_1/dy_z|_{y_z=0}$ at both $p_T=5$ and $6$ GeV$/c$ are found to be about - 4$\times10^{-3}$ for both charm and bottom quarks
within the statistical uncertainty of the calculation. 
This can be expected because, as shown before, the Lorentz force effect is similar at $p_T>m$, and the mass effect will thus 
be negligible at high $p_T$.

\begin{figure}[h]
\centering
\includegraphics[width=1\linewidth]{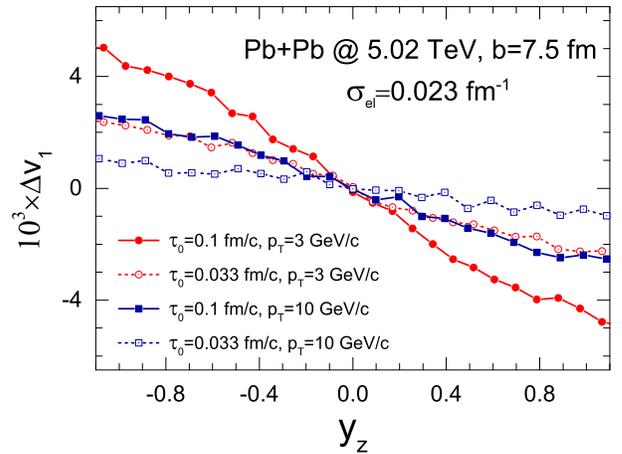}
\caption{(Color online) The comparison between  $\Delta v_1$ vs $y_z$ of  mesons fragmented by heavy quarks with the formation time set as 0.1 fm$/c$ and 0.033 fm$/c$ at $p_T=$ 3 and 10 GeV$/c$.
}
\label{fig:time}
\end{figure}
Finally we study the effect of the formation time on $\Delta v_1$. The formation time of heavy quarks is given by the pair production process that is approximated as $1/2m$, and we thus vary $\tau_0$ from 0.1 fm$/c$ to 0.033 fm$/c$ to see how it affects $\Delta v_1$. 
The numerical results at $p_T=$ 3 GeV$/c$ and 10 GeV$/c$ are shown in Fig.~\ref{fig:time}, where the slopes $d\Delta v_1/dy_z|_{y_z=0}$ at $p_T=3$ GeV$/c$ are found to be -5.7$\times10^{-3}$ with charm formation time and -3.1$\times10^{-3}$ with bottom formation time and -2.9$\times10^{-3}$ with charm formation time and -1.2$\times10^{-3}$ with bottom formation time at $p_T=10$ GeV$/c$.
The change is seen to be surprisingly large with such a small change in the formation time. The results can be understood
again by Eq. (\ref{slope}), or Eq. (\ref{scaling}) where $tB_y$ at $\tau_1$ does not change, but it changes significantly at $\tau_0=0.1$ and 0.033 fm$/c$, see the blue dash-dotted line in Fig.~\ref{fig:tBy}. 
Our Eq. (\ref{scaling}) shows that more generally 
$\Delta v_1$ is sensitive to $\tau_0$ only when the  difference of $tB_y$ at $\tau_1$ and $\tau_0$ is dominated by $tB_y$ at $\tau_0$; 
if the time dependence of the magnetic field is such that 
$tB_y$ at $\tau_1$ dominates over its value at $\tau_0$, $\Delta v_1$ should not change significantly by varying $\tau_0$.  

\begin{figure}[h]
\centering
\includegraphics[width=1\linewidth]{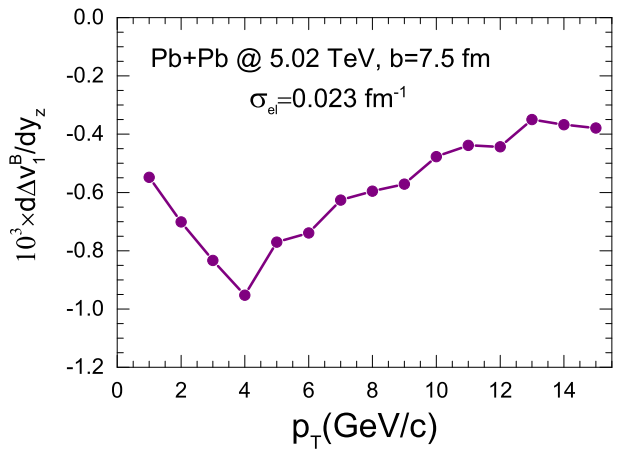}
\caption{(Color online) 
$d\Delta v_1/dy_z|_{y_z=0}$ of B mesons generated by e.m. fields with the medium conductivity $\sigma_{el}=$0.023fm$^{-1}$ in 5.02 TeV Pb+Pb collisions at $b=$ 7.5 fm.
}
\label{fig:Bmeson}
\end{figure}
In Fig. \ref{fig:Bmeson} we show the results for  $d\Delta v_1^B/dy_z|_{y_z=0}$ for the case of B mesons at $\sigma_{el}=0.023 fm^{-1}$.
We can see the slope of the splitting is about a factor of six smaller than the charm case, see Fig. \ref{fig:charm}. 
Such a factor arises from a factor of 2 from the charge,
about a factor of 2 from the smaller formation time, and roughly about a factor 1.5 from the smaller slope of the bottom spectrum at 
$p_T \simeq 5-10 \rm \,GeV/c$

\subsection{The correlation between charmed mesons and leptons from the decay of $Z^0$}
As  shown in the section above, 
the charm and bottoms are so different that we may not be
able to determine safely whether the experimental measurement of $\Delta v_1$ for B and D
have sole e.m. fields origin. Moreover, to extract both $\alpha$ and $\beta$ from experimental data one needs 
to use the scaling $\alpha\frac{-\partial \ln f}{\partial p_T}+\frac{2\alpha-\beta}{p_T}$ to fit the data at high $p_T$ 
where the interaction with QGP plays a negligible role on the charge dependent flow observables; 
However, due to $\frac{-\partial \ln f}{\partial p_T}\propto \frac{1}{p_T}$ at high $p_T$ because of the power law decay of the spectra of  quarks at high $p_T$, it is hard to identify $\alpha$ and $\beta$ separately, because the last is also $\propto \frac{p_T}{m^2_T}\sim \frac{1}{p_T}$.

On the other hand, in Ref~\cite{Sun:2020wkg}, leptons from $Z^0$ decay are found to be an excellent probe of e.m. fields based on the following reasons: (i) Leptons weakly interact with QGP and do not have complex hadronization mechanisms as heavy and light quarks so as to be a cleaner probe; (ii) Leptons from the decay of $Z^0$ share a similar formation time $\tau_0=1/2.5$ GeV$^{-1}$ as charm quarks, so that all the coefficients $a_n$, $b_n$ and $c_n$ 
of leptons can be approximated as 1.5 times those of charm quarks, because they experience the same $\Delta(t B_y(t))$ as charm.
This makes an important difference with respect to bottom quarks because it resets the uncertainty coming from the difference in the 
formation time, discussed above.
A test of this should be a strong probe of e.m. fields; (iii) Leptons from the decay of $Z^0$ have a peculiar spectra so that it is easier to identify 
$\alpha, \beta$ coefficients; (iv) Since the Lorentz force becomes same at $p_T\gg m$, a measurement of constant $\alpha$ at such high $p_T$ 
by leptons from $Z^0$ decay should also be a  strong probe of e.m. field. 

In this section, we will study how  $\Delta v_1$ of leptons correlates
 with charm quarks, and see if it fulfills our expectation, through numerical simulations for 5.02 TeV Pb+Pb collisions at $b=7.5$ fm. 
 We will also show how the general formula (\ref{distribution}) instructs us to make general predictions for the leptons.

The spectra of leptons is generated by decaying $Z^0$ into lepton pairs, where we first obtain the momentum distribution of $Z^0$ by fitting the experimental measurements~\cite{Chatrchyan:2014csa,Khachatryan:2015pzs}:
\begin{eqnarray}
dN/d^2p_Tdy_z=f(\mathbf{p_T},y_z)\propto 10^{-ap_T^{n}}e^{-\frac{y_z^2}{2 \Delta_l^2}}.
\end{eqnarray}
The parameters $a=0.6896$, $n=0.4283$ and $\Delta_l=3.034$ are found to give quite a good description of $p_T$  and $y_z$  dependence of $Z^0$ in 5.02 TeV Pb+Pb collisions~\cite{Sun:2020wkg}. From the spectra of $Z^0$, the spectra of lepton can be obtained: 
\begin{eqnarray}
&&\frac{dN_l}{d^2p_{T1}dy_1}=\frac{E_1dN_l}{d^3p_1}\nonumber
\\&&=\frac{\Gamma_l}{\Gamma_{tot}}\int \frac{d^3p_2}{E_2} \frac{d^3 p_3}{E_3} \delta^4 (p_1+p_2-p_3) \frac{m_{Z^0}}{4\pi p_f}  f(\mathbf{p}_{T3},y_3)\nonumber
\\&&= \frac{\Gamma_l}{\Gamma_{tot}} \frac{m_{Z^0}}{2\pi p_f}\int d^2p_{T2}\frac{f(\mathbf{p}_{T1}+\mathbf{p}_{T2},y_3)}{|2m_{T1}m_{T_2}\sinh(y_1-y_2)|},
\label{decay}
\end{eqnarray}
where $y_2$ and $y_3$ are given by the energy conservation and have two sets of solution:
\begin{eqnarray}
&&\sqrt{m_{Z^0}^2+(\mathbf{p}_{T1}+\mathbf{p}_{T2})^2}\cosh y_3\nonumber
\\&&=m_{T1}\rm{cosh}y_1+m_{T2}\rm{cosh}y_2,
\\&&m_1^2+m_2^2-M_{Z^0}^2+2m_{T1}m_{T2}\cosh(y_1-y_2)\nonumber
\\&&=2p_{T1}p_{T2}\cos \phi.
\end{eqnarray}
In the above, $\Gamma_l/\Gamma_{tot}$ is the branching ratio, $p_f$ is the magnitude of momentum of each lepton in COM frame of $Z^0$ ($2\sqrt{p_f^2+m_1^2}=m_{Z^0}$), $m_1=m_2$ are the mass of lepton pairs, and $\phi$ is the angle between the transverse momentum of lepton pairs.

The transverse coordinate of $Z^0$ is given by the binary collisions of colliding nuclei, and the formation time $t$ and longitudinal coordinate $z$ are given by $t=\tau_{Z^0}\cosh y_z$ and $z=\tau_{Z^0}\sinh y_z$ with $\tau_{Z^0}=1/m_{Z^0}=0.0022$ fm$/c$. Finally the spacetime coordinate of produced leptons is given by their mother $Z^0$ that moves in a straight line with a decay time having  a distribution $\rho(\Delta t)\propto e^{-\frac{\Gamma_{tot}\Delta t}{\gamma_v}}$ with $\Gamma_{tot}=2.495$ GeV and $\gamma_v$ being the Lorentz contraction factor.

\begin{figure}[h]
\centering
\includegraphics[width=1\linewidth]{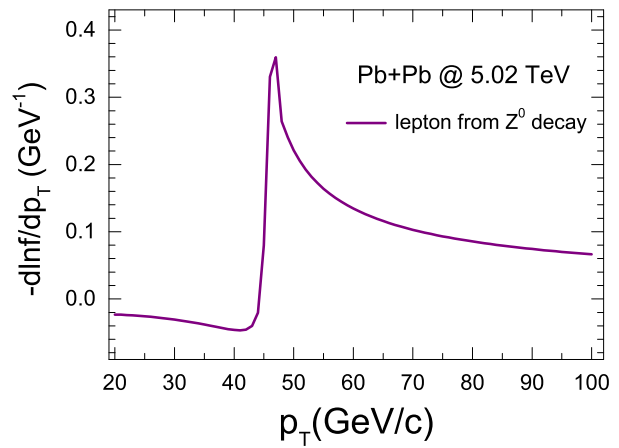}
\caption{(Color online)  $-\frac{\partial \ln f}{\partial p_T}$ of leptons from the decay of $Z^0$ in 5.02 TeV Pb+Pb collisions.
}
\label{fig:muonspectra}
\end{figure}

In Fig.~\ref{fig:muonspectra}, we show $-\frac{\partial \ln f}{\partial p_T}$ of leptons from the decay of $Z^0$ deduced by Eq. (\ref{decay}). It is seen that $-\frac{\partial \ln f}{\partial p_T}$ is negative at $p_T<m_{Z^0}/2=45$ GeV$/c$, and it jumps to a large and positive value above 45 GeV$/c$ due to the kinematic effect. This peculiar spectra should imprint a signature in the spectra ratio and $\Delta v_n$ of positively and negatively leptons inspired 
by the general formula in Eq. (\ref{distribution}).

\begin{figure}[h]
\centering
\includegraphics[width=1\linewidth]{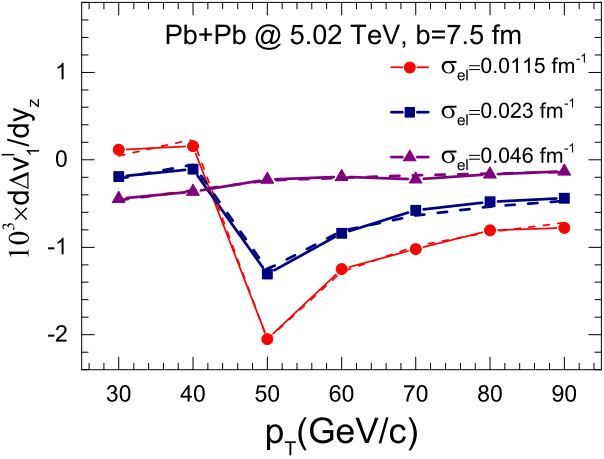}
\caption{(Color online)  $d\Delta v_1^l/dy_z|_{y_z=0}$ of lepton pairs generated by e.m. fields with the medium conductivity $\sigma_{el}=$ 0.0115, 0.023 and 0.46 fm$^{-1}$ in 5.02 TeV Pb+Pb collisions at $b=$ 7.5 fm. The dashed lines are the fittings with $-\alpha\frac{\partial \ln f_l}{\partial p_T}+(2\alpha-\beta)\frac{p_T}{m_T^2}$.
}
\label{fig:muon}
\end{figure}

The results of  $d\Delta v_1^l/dy_z|_{y_z=0}$ of lepton pairs  in 5.02 TeV Pb+Pb collisions at $b=$ 7.5 fm, which is generated by e.m. fields with  $\sigma_{el}=$ 0.0115, 0.023 and 0.46 fm$^{-1}$, are shown in Fig.~\ref{fig:muon}. 
In general, there is a sudden drop of the slope at $p_T=m_{Z^0}/2$ with a peak structure that is driven by $\frac{\partial \ln f_l}{\partial p_T}$,
according to the Eq. (\ref{scaling}) $-\alpha\frac{\partial \ln f_l}{\partial p_T}+(2\alpha-\beta)\frac{p_T}{m_T^2}$.
However for the conductivity  $\sigma_{el}=0.046$ fm$^{-1}$ such a peak structure disappears, again Eq. (\ref{scaling}) allows to understand it; 
in fact in this case $\tau_0 B_y(\tau_0) \simeq \tau_1 B_y(\tau_1)$  and hence the $\alpha$ coefficient, multiplying $\frac{\partial \ln f_l}{\partial p_T}$,  
becomes quite small.
The fittings with our formula are shown by the dashed lines in Fig.~\ref{fig:muon} agree with the numerical results quite well. 
The $\alpha$ factor  for $\sigma_{el}=0.046 \, \rm fm^{-1}$ becomes about 50 times than the case for $\sigma_{el}=0.0115 \, \rm fm^{-1}$, essentially
because the difference in $tB_y(t)$ at formation time and escape time are nearly equal.
The $\alpha$ ratio of leptons from $Z^0$ decay and charm quarks, which is about $4.7/3.6=1.3$ and $8.7/6.3=1.38$ for electrical conductivity 0.023 fm$^{-1}$ and 0.0115 fm$^{-1}$ separately, is close to their charge ratio (implying a quite
similar value of the K factor), which confirms the strong correlation
between the $\Delta v_1$ of charm and leptons,
once subtracting the impact of the very different $p_T$ slope of the spectrum.

\begin{figure}[h]
\centering
\includegraphics[width=1\linewidth]{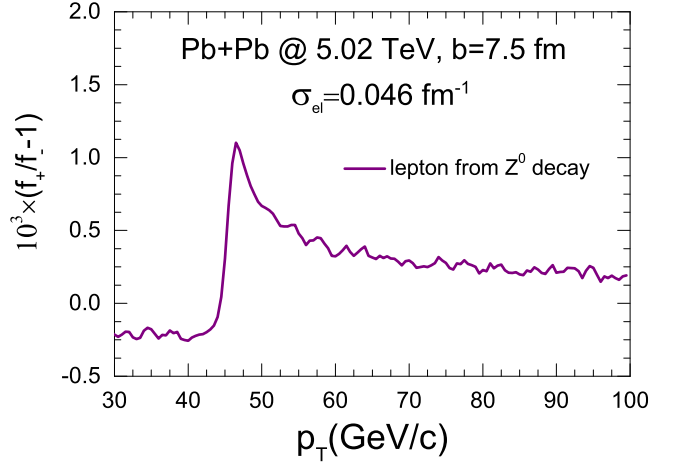}
\caption{(Color online)  The ratio of the spectra of positively and negatively charged leptons in midrapidity $|y_z|<0.5$ from $Z^0$ decay as a function of $p_T$ with the medium conductivity $\sigma_{el}=$ 0.046 fm$^{-1}$ in 5.02 TeV Pb+Pb collisions at $b=$ 7.5 fm.
}
\label{fig:ratio}
\end{figure}
We close our analysis with a final consideration.
As shown in Eq. (\ref{distribution}), the peculiar spectra of leptons from $Z^0$ decay should leave fingerprints in both the spectra $f(p_T,\phi, y_z)$
and $\Delta v_n$ of positively and negatively charged particles, as long as $a_n$, $b_n$ and $c_n$ are non-zero. For example, with the help of the first two lines of Eq. (\ref{distribution}) and by knowing that $a_n$, $b_n$ and $c_n$ do not depend on $p_T$ when $p_T\gg m$, their spectra after the effects of e.m. fields become according to Eq. (\ref{distribution}):
\begin{eqnarray}
f^{'}|_{y_z=0}=f\left[1-(a_1+b_1)\frac{\partial \ln f}{\partial p_T}-\frac{2}{p_T}\frac{\partial c_0}{\partial y_z}\right]|_{y_z=0}.
\end{eqnarray}
where $a_1$ and $b_1$ become non-zero when $E_x$ and $E_y$ at $\eta_s=0$ have non-zero $\cos \phi_x$ and $\sin \phi_x$ terms respectively.
We thus also studied the ratio of the spectra of positively and negatively leptons from $Z^0$ in 5.02 TeV collisions at $b=7.5$ fm using 
$\sigma_{el}=0.046$ fm$^{-1}$, and the results are shown in Fig.~\ref{fig:ratio}. It is seen that the ratio, $f_+/f_- -1$, is driven by the term
$-\frac{\partial \ln f}{\partial p_T}$ of leptons shown in Fig.~\ref{fig:muonspectra}, though the ratio is very close to 1 (deviated by 10$^{-3}$), 
which means that $a_1$ and $b_1$ are non-zero. 
On the other hand, if one looks at the spatial configurations of $E_x$ and $E_y$ \cite{Deng:2012pc,PhysRevC.94.044903,Gursoy:2018yai}, 
one should immediately identify large dipole moments and conclude that they should be non-zero.

\section{Conclusions and Discussions}
In this study we have obtained the general formula of the charge dependent flow observables  generated by e.m. fields, 
which has a simple form $\Delta v_n(p_T,y_z)=-\frac{\Delta d_n(y_z)}{2}\frac{\partial \ln f}{\partial p_T}-\frac{\Delta e_n(y_z)}{2p_T}$ at high $p_T$ according to Eq. (\ref{flow}) where the specific impact of strong interactions is subdominant.
An experimental check of the $p_T$ pattern it predicts for the splitting of matter/anti-matter $\Delta v_1$
and the correlations between the charm meson directed flow and the one of leptons from $Z^0$ decay (yet to be measured)
would provide a strong probe of e.m. fields. 
The coefficients in the formula have a direct relation to the expansion of e.m. fields in $\phi_x$, and the Lorentz force in the longitudinal direction 
contributes also to the charge dependent flow observables that measure the anisotropy in transverse momenta. 
Moreover the strength of our formula is to trace back the sign and the strength of the splitting $\Delta v_1$ to time dependence
of the magnetic field $B_y(t)$, in the center of the colliding system, including also the effect of the electric field $E_x$ generated by the Faraday's law.
This has been obtained
under the approximation that the space gradients of the electromagnetic field can be discarded within the core of the QGP fireball
created in AA collisions. The formula derived allows to understand that the intial strength of the magnetic field does not determine the sign and/or the strength
of $\Delta v_1$. Furthermore clarify also the relation between the $\Delta v_1$ of charm and bottom,
and moreover the one with the leptons from $Z^0$ decay that appears quite different due to the very different $p_T$ dependence of the spectra.
To confirm the validity of the formula, we have compared it to the realistic numerical simulation 
in a relativistic Langevin approach, finding a very good agreement between them. By comparing the numerical results between charm and bottom quarks, we show how the formula serves as a powerful tool to understand the results. Finally, we pointed out the strong correlation between the coefficients $a_n$, $b_n$ and $c_n$ of charm quarks and leptons from $Z^0$ decay, where $a_n$, $b_n$ and $c_n$ are the respective harmonic expansions of the  mean variations of the three momenta due to e.m. fields as seen by Eq. (\ref{shifts}), which is believed to hold for any configuration of e.m. fields.

The present study  utilizes the Peterson fragmentation  converting heavy quarks into heavy mesons, and so the shape does not modify much from quarks, 
that can not be directly probed, to mesons. However, the hadronization mechanism by coalescence plus fragmentation \cite{PhysRevC.73.034913,Scardina:2017ipo,Dong:2019unq} may modify it 
quantitatively since the coalescence model combines one heavy quark with light quarks of different $p_T$ into mesons by a non random selection
of the bulk matter along the hypersurface of hadronization. 
However, if one looks at the high $p_T$ behavior discussed along the present work, 
the modification should be small since the coalescence contribution significantly decreases with $p_T$.

The present paper is presenting a test of a pocket formula for $\Delta v_1$ by comparison to realistic simulations of AA collisions
but under e.m. space-time profile coming from the assumption of the existence of an equilibrated QGP matter at constant $\sigma_{el}$.
However the analysis could be extended also to other e.m. profiles that currently under consideration for example in the study of the chiral magnetic effect
\cite{Huang:2017tsq}.
Our study shows that the value of the magnetic field at the very early time, $t< 0.1 \rm\, fm/c$, can significantly modify the relation between the
$\Delta v_1$ of D and B mesons due to the relevance of the value of $tB_y(t)$ at the the formation time of charm and bottom quarks.

\section*{ACKNOWLEDGEMENTS}
The work of  Y.S. is supported by a INFN post-doc fellowship within the national SIM project.
S.P. and V.G. acknowledge the support of INFN-SIM national project and linea di intervento 2 for HQCDyn at DFA-Unict. S.P. acknowledge the funding from UniCT under 'Linea di intervento 2' (HQsmall Grant).

\bibliography{ref.bib}
\end{document}